\documentclass[twocolumn,amsmath,amssymb,floatfix]{revtex4}

\usepackage{bm}
\usepackage{amsmath}
\usepackage{epsf}

\newcommand{\beq}{\begin{equation}}
\newcommand{\eeq}{\end{equation}}
\newcommand{\beqa}{\begin{eqnarray}}
\newcommand{\eeqa}{\end{eqnarray}}
\newcommand{\data}[1]{{\bf d}_{#1}}
\newcommand{\proj}[1]{{\bf P}_{#1}}
\newcommand{\weight}[1]{{\bf R}_{#1}}
\newcommand{\chol}[1]{{\bf L}_{#1}}
\newcommand{\esignal}[1]{\hat{\bf s}_{#1}}
\newcommand{\signal}[1]{{\bf s}_{#1}}
\newcommand{\iden}{{\bf I}}
\newcommand{\noise}[1]{{\bf n}_{#1}}
\newcommand{\eigen}{{\bf e}}
\newcommand{\csignal}[1]{{\bf S}_{#1}}
\newcommand{\cnoise}[1]{{\bf N}_{#1}}
\newcommand{\dela}{{\Delta}}
\newcommand{\dD}{{\delta\!D}}
\newcommand{\bn}{\hat{\bf n}} 

\newcommand{\etal}{{\it et al. }}

\newcommand{\ApJL}{Astrophys. J Lett.}
\newcommand{\ApJ}{Astrophys. J}
\newcommand{\PRL}{Phys. Rev. Lett.}
\newcommand{\PRD}{Phys. Rev. D}
\newcommand{\MNRAS}{Mon. Not. Roy. Astr. Soc.}
\newcommand{\ARAA}{Ann. Rev. Astron. Astrophys.}
\newcommand{\AsAs}{Astron. Astrophys.}
\newcommand{\amp}{\& }
\newcommand{\aut}[2]{{#2.\ #1,}}
\newcommand{\laut}[2]{{#2.\ #1,}}
\newcommand{\refs}[6]{#2, {\bf #3},  {#4} (#5).}
\newcommand{\mrefs}[6]{#2, {\bf #3},  {#4} (#5);}
\newcommand{\urefs}[5]{#2, #3, #4 (#5).}
\newcommand{\murefs}[5]{#2, #3, #4 (#5);}
\newcommand{\mybib}[2]{\bibitem{#2}}

\begin{document}

\title{Three-Dimensional Mapping of the Dark Matter}
\author{Wayne Hu$^{1,2}$ \& Charles R. Keeton$^{2}$\footnote{Hubble Fellow}}
\affiliation{
{}$^{1}$Center for Cosmological Physics, University of Chicago, Chicago IL 60637 \\
{}$^{2}$Department of Astronomy and Astrophysics, University of Chicago, Chicago IL 60637
}

\begin{abstract}
We study the prospects for three-dimensional mapping of the dark
matter to high redshift through the shearing of 
faint galaxies images at multiple distances by gravitational lensing.   
Such maps could provide invaluable information on the nature of the
dark energy and dark matter.  While in principle 
well-posed, mapping by direct inversion introduces exceedingly large,
but usefully correlated noise into the reconstruction.  By carefully
propagating the noise covariance, we show that lensing
contains substantial information, both direct and statistical, on the 
large-scale radial evolution of the density field.  This information
can be efficiently distilled into low-order signal-to-noise 
eigenmodes which may be used to compress the data by over an order of
magnitude.   Such compression will be useful for the
statistical analysis of future large data sets.
The reconstructed map also contains useful information on the localization of 
individual massive dark matter halos, and hence the dark energy
from halo number counts,  but its extraction depends strongly on 
prior assumptions.  We outline a procedure for maximum 
entropy and point-source regularization of the maps that can identify
alternate reconstructions.
\end{abstract}
\maketitle

\section{Introduction}

Embedded within the evolution of the three-dimensional distribution 
of the dark matter lies a wealth of information on the nature of 
the dark energy and dark matter in the universe.  The growth of
its clustering in volumes associated with redshift, along with the 
abundance of discrete dark matter clumps or halos which it controls,
is one of our most direct
probes of the expansion history
(e.g.\ \cite{deprobes}).   
It is certainly the one
that is best understood from the theoretical standpoint.

Unfortunately, most probes of structure in the low redshift universe
rely on luminous matter, e.g.\ galaxies and clusters of galaxies,
as tracers of dark matter distribution.
Such probes are subject to significant uncertainties in the physical
processes that govern the formation and evolution of the tracers.
The only exception is the image distortion from
gravitational lensing of distant objects 
by the dark matter.  On large scales and for small distortions, 
this is known as weak gravitational lensing \cite{weak,weaklss}.
The distortion of faint galaxy images by the large-scale structure 
of the universe has now been detected with high significance 
by several experimental groups \cite{weakdet}.  

A fundamental obstacle for weak lensing studies of the matter
distribution is that the technique is inherently two-dimensional.
All of the matter along the line-of-sight to a distant source
contributes to lensing and so the distortion reflects 
a two-dimensional projection of the dark matter.  
Unfortunately then the evolution of structure is hidden in the
missing radial dimension.
This limitation can in principle be overcome by a tomographic
reconstruction of the three-dimensional distribution from
sources spanning a range of distances or redshifts.  

Under fairly restrictive assumptions, this tomographic technique 
has been applied to lensing data to
localize the halo associated with a cluster of galaxies \cite{Witetal01} and
validated by follow-up studies. 
The critical assumption is that the lensing mass be a single halo, 
well localized in redshift.
Taylor \cite{Tay01} has recently shown that 
these and other restrictions are unnecessary
{\it in principle}. 
In the absence of noise, tomographic mapping of the dark matter
is a well-posed problem.  In this paper, we study the feasibility
of reconstructing three-dimensional dark matter maps in the
presence of noise.  We will show that a careful accounting
of the noise and in particular its {\it covariance\/} across the map
is essential for extracting information from the map. 

Tomography also presents a severe data analysis challenge, similar
to but potentially far worse than that facing cosmic microwave
background (CMB) experiments.   A full weak lensing data set will
have a two-component, two-dimensional megapixel map for each of
ten or more source redshift slices.
We also study how techniques developed for the CMB and galaxy 
redshift surveys may be applied to
compress these data to a more manageable size.  

The outline of the paper is as follows.  We begin in 
\S~\ref{sec:formalism} with a brief review of mapmaking
techniques and apply them to the two-dimensional lensing observables keeping
careful track of the propagation of measurement errors.  
We use these techniques to reconstruct the three-dimensional
distribution of the dark matter in \S~\ref{sec:los}.
Although the reconstruction is extremely noisy for any
realistic situation, the noise has very particular properties
that are absent in the signal.  
This fact is used to regularize the solution and
radically compress the data in the large-scale structure 
regime in \S~\ref{sec:lss}, and in the individual dark matter
halo regime in \S~\ref{sec:halos}.   We discuss these
results in \S~\ref{sec:discussion}.

\vfill

\section{Formalism}
\label{sec:formalism}

We begin by briefly reviewing general mapmaking techniques 
in \S~\ref{sec:mapmaking}, establishing notation 
used throughout the paper.  These techniques are then applied to the
two-dimensional weak lensing shear 
in \S~\ref{sec:observables} for the reconstruction of
convergence maps.  The latter follows the well-known Kaiser \& Squires
\cite{KaiSqu93} algorithm but we pay special attention to the
propagation of noise into the convergence reconstruction as that
will play a central role in the three-dimensional mapping that follows.  
In \S~\ref{sec:sourceplane}, we discuss the generalization to multiple
source planes.

\subsection{Mapmaking}
\label{sec:mapmaking}

Mapmaking can be formulated in terms of the
general inverse problem \cite{Teg97}, where we seek an
estimate $\esignal{a}$ of a signal vector $\signal{a}$ from
a data vector $\data{b}$
that is a linear projection $\proj{ba}$ 
of the signal plus measurement noise $\noise{b}$,
\begin{equation}
\data{b} = \proj{ba} \signal{a} + \noise{b}\,.
\label{eqn:mapmaking}
\end{equation}
The projection matrix $\proj{ba}$ has dimensions ($n_b$,$n_a$),
the number of elements in the data and signal vectors respectively,
and hence need not be square.  Here and below subscripts are labels
and not elements of the vectors and matrices; elements will be
denoted by $[\proj{ba}]_{ij}$.  We assume that
both the signal and the noise have zero mean 
$\langle \signal{a} \rangle = \langle \noise{b} \rangle = 0$,
that the signal and noise are uncorrelated,
$\langle \signal{a} \noise{b}^t \rangle = 0$,
and that the noise covariance is known,
\begin{equation}
\cnoise{bb} \equiv \langle \noise{b} \noise{b}^t \rangle \,.
\end{equation}
The statistical properties of the signal,
\begin{eqnarray}
\csignal{aa} &\equiv& \langle \signal{a} \signal{a}^t \rangle \,,
\end{eqnarray}
may or may not be known.

The estimated signal $\esignal{a}$ is that which minimizes 
\begin{equation}
\chi^2 + H\,,
\label{eqn:minimize}
\end{equation}
where
\begin{equation}
\chi^2 = 
(\data{b} - \proj{ba} \esignal{a})^t 
\cnoise{bb}^{-1}
(\data{b} - \proj{ba} \esignal{a})\,. 
\end{equation}
and the penalty function $H$ is a set of constraints and/or
a regularization to choose among degenerate solutions.

Minimizing $\chi^2$ (with $H=0$) returns the linear estimator
\begin{equation}
\esignal{a} = \weight{ab} \data{b} \,,
\label{eqn:estimator}
\end{equation}
where
\begin{equation}
\weight{ab} = [\proj{ba}^t \cnoise{bb}^{-1} \proj{ba}]^{-1} 
	\proj{ba}^t \cnoise{bb}^{-1}\,,
\label{eqn:minimumchi2}
\end{equation}
and this simple reconstruction is well-posed as long as the
product in square brackets is invertible.  If $\proj{ba}$
itself is invertible then $\weight{ab} = \proj{ba}^{-1}$ and
the estimator becomes independent of both the signal and the
noise.  The errors in the reconstruction,
\begin{equation}
\esignal{a} - \signal{a} = [\weight{ab} \proj{ba} - \iden ] \signal{a} +
			   \weight{ab} \noise{b}\,,
\end{equation}
imply a new noise covariance
\begin{eqnarray}
\cnoise{aa} &\equiv& 
	\langle (\esignal{a}-\signal{a})(\esignal{a}-\signal{a})^t\rangle
\nonumber\\
&=& [\proj{ba}^t \cnoise{bb}^{-1} \proj{ba}]^{-1}\,, 
\label{eqn:estimatornoise}
\end{eqnarray}
which is independent of the signal.
Note that minimizing $\chi^2$ is not the same as minimizing the
reconstruction noise $\cnoise{aa}$ since reconstruction
errors are not penalized by $\chi^2$ in the noise-dominated regime. In fact,
it minimizes $\cnoise{aa}$ subject to
the constraint $\weight{ab} \proj{b a} = \iden$ \cite{Teg97}. 

For a noisy reconstruction, prior knowledge of the statistical
properties of the signal can be used as to set a penalty function
for a new estimator of the signal
\begin{equation}
H = \esignal{w}^t \csignal{aa}^{-1} \esignal{w}\,.
\end{equation}
Minimization of Eqn.~(\ref{eqn:minimize}) then returns the
Wiener filtered estimate of the signal 
$\esignal{w} = \weight{wb}\data{b}$, where
\begin{eqnarray}
\weight{wb} &=& 
[\csignal{aa}^{-1} + \proj{ba}^t \cnoise{bb}^{-1} \proj{ba} ]^{-1}
\proj{ba}^t \cnoise{bb}^{-1} \nonumber\\
	    &=&
\csignal{aa} \proj{ba}^t [ \proj{ba} \csignal{aa} \proj{ba}^t +
			   \cnoise{bb}]^{-1}\,,
\label{eqn:wiener}
\end{eqnarray}
which has the heuristic form of signal/(signal+noise) and so
suppresses the signal in  the noise-dominated regime. 
The estimator $\esignal{w}$ has the noise properties
\begin{eqnarray}
\cnoise{ww} 
	  &=& 
	[\weight{wb}\proj{ba} - \iden ] \csignal{aa}
	[\weight{wb}\proj{ba} - \iden ]^t  \nonumber\\
&& \quad
	+ \weight{wb} \cnoise{bb} \weight{wb}^t \,.
\label{eqn:wienernoise}
\end{eqnarray}
The Wiener estimator may alternately be derived as that which
minimizes $\cnoise{ww}$ \cite{Teg97}.  

For a well-posed inverse
problem the minimum-$\chi^2$ and Wiener reconstructions are
related by an invertible operation.  
We can view the result of the former as providing 
a new data vector $\data{a} = \esignal{a}$ having noise
$\noise{a}$ with a 
covariance $\cnoise{aa}$.    Then with the model of the
data $\data{a} =\iden\ \signal{w} + \noise{a}$,  
Eqn.~(\ref{eqn:wiener}) for the Wiener filter returns
\begin{equation}
\esignal{w} = \weight{wa}\data{a} = \weight{wa}\esignal{a}\,,
\label{eqn:wienerop}
\end{equation} where
\begin{equation}
\weight{wa} = \csignal{aa} [\csignal{aa} + \cnoise{aa}]^{-1}\,.
\end{equation}
Since this matrix is invertible, the two reconstructions are 
formally equivalent.

Because the minimum-$\chi^2$ method does not require prior
knowledge, we choose it as the primary mapping technique if
the inverse problem is well-posed.
Alternatively, for an ill-posed inverse problem, where the
matrix $\proj{ba}^t \cnoise{bb}^{-1} \proj{ba}$ is not
invertible, Wiener filtering can serve as the primary
technique.

As the Wiener example implies, secondary processing operations on the
primary map can be viewed as 
simply another round of mapmaking.
Any linear operation that is invertible will retain the same information content
as the original so long as the noise covariance is properly propagated.
We will use this technique to go from the observed lensing shear to
the convergence to the density field and finally to the Wiener, 
signal-to-noise, or point source filtered density fields.

\subsection{Lensing Observables}
\label{sec:observables}

The distortion of images due to weak gravitational lensing is described
by the Jacobian matrix of the mapping between the two-dimensional
source and image 
planes (e.g.\ \cite{BarSch01})
\begin{eqnarray}
\label{A}
{\bf A} =
\left(
\begin{array}{cc}
1-\kappa-\gamma_1 & -\gamma_2\\
-\gamma_2 & 1-\kappa+\gamma_1
\end{array}
\right) \, ,
\end{eqnarray}
where all components are functions of position on the sky $\bn$.

The galaxy ellipticities form a noisy estimator of the
shear components $(\gamma_1,\gamma_2)$ which are in turn
related to the convergence $\kappa$ by (e.g.\ \cite{BarSch01})
\begin{equation}
[\gamma_1 \pm i \gamma_2](\bn) = -\frac{1}{\pi} \int d\bn ' 
\frac{e^{\pm 2i\phi}}{\theta^2} \kappa(\bn')\,,
\label{eqn:shearkappa}
\end{equation}
where $\theta \equiv |\bn - \bn'|$ denotes the length of
the pixel separation vector, and $\phi$ denotes its
azimuthal angle in the coordinate system that
defines the shear components.  
By constructing a map of $\kappa$ from the shear data, one
compresses the data set by a factor of two and also transforms
the data into a form that is more conveniently related to the
three-dimensional density field (see \S~\ref{sec:los}).

Discretizing the sky into pixels returns Eqn.~(\ref{eqn:shearkappa})
in the form of the general mapmaking
problem of Eqn.~(\ref{eqn:mapmaking}), where the signal $\signal{\kappa}$
has been linearly projected onto the data space 
with measurement noise added to form the data vector,
$\data{\gamma}$.
Explicitly, if one orders the data vector as 
\begin{equation}
\data{\gamma} = \{ \gamma_1(\bn_1), \gamma_2(\bn_2); \ldots
; \gamma_1(\bn_{n_{\rm pix}}),\gamma_2(\bn_{n_{\rm pix}})\} \,,
\label{eqn:sheardata}
\end{equation}
the model becomes
\begin{equation}
\data{\gamma} = \proj{\gamma\kappa} \signal{\kappa} + \noise{\gamma}\,,
\label{eqn:shearmodel}
\end{equation}
with the projection matrix 
\begin{eqnarray}
\left[\proj{\gamma\kappa}\right]_{(2i-1)j} &=& -\frac{A_j}{\pi} 
	\frac{\cos 2\phi_{ij}}{\theta^2_{ij}} 
		\nonumber\,,\\
\left[\proj{\gamma\kappa} \right]_{(2i)j}   &=&   -\frac{A_j}{\pi}   
	\frac{\sin 2\phi_{ij}}{\theta^2_{ij}} \,,
\end{eqnarray}
where the indices run over the pixels of area $A_j$ and the angles are
defined as averages over the pixel. 

If the noise in the
shear data is dominated by the intrinsic ellipticity of galaxies,
it is given by
\begin{equation}
[\cnoise{\gamma\gamma}]_{ij} = [\iden]_{ij} \frac{\gamma_{\rm rms}^2}
{[{\bf n}_{\rm gal}]_i}\,,
\label{eqn:shearnoise}
\end{equation}
where ${\bf n}_{\rm gal}$ is a vector containing 
the number of galaxies per pixel. Here $\gamma_{\rm rms}$ is
the rms error from intrinsic ellipticities and measurement errors
per galaxy.  The intrinsic alignment of galaxies on small scales
is a potential source of correlated noise \cite{intrinsic}.  
While we neglect its currently uncertain contribution 
here, the framework we establish can handle any source of
noise provided its covariance is known.

We have implicitly assumed here a single set of shear data per pixel. 
If the full data set includes multiple observations of the shear of
varying quality or even simply the unbinned individual galaxy estimates
themselves,  one merely extends the data vector and the mapmaking
algorithm combines them with the appropriate noise weighting.
Mapmaking can also test the validity of 
Eqn.~(\ref{eqn:shearkappa}),
or more properly the data model of
Eqn.~(\ref{eqn:shearmodel}), through a reconstruction of
the complementary ``B-mode'' map.
This procedure amounts to 
multiplying the kernel in Eqn.~(\ref{eqn:shearkappa}) by $i$ 
corresponding to a rotation of the shear vectors by $45^\circ$.
The reconstructed map should be consistent with noise.  

The estimator of $\kappa$ and its noise properties then follow from
Eqns.~(\ref{eqn:estimator}) and (\ref{eqn:estimatornoise}) aside from
two subtleties.  The first subtlety involves the so-called
mass-sheet degeneracy: adding a constant to the $\kappa$ signal in
Eqn.~(\ref{eqn:shearkappa}) yields no effect in the shear data.
There is then a singular value associated with the inversion in
Eqn.~(\ref{eqn:minimumchi2}).  In general, one way to handle
unconstrained modes is to add them to the data vector and assign
them zero value but a large noise variance \cite{unconst}.  
The covariance matrix of Eqn.~(\ref{eqn:estimatornoise}) then
properly accounts for the lack of information on these modes.
For the mass-sheet degeneracy, one appends:
a zero to the data vector in Eqn.~(\ref{eqn:sheardata});
a row to the projection matrix with uniform elements $1/n_{\rm pix}$;
and a diagonal entry to the noise matrix Eqn.~(\ref{eqn:shearnoise})
with a value substantially greater than the variance of $\kappa$
smoothed across the field size in any reasonable cosmology.

This same procedure applies to the second subtlety.
Since Eqn.~(\ref{eqn:shearkappa}) represents a convolution, the discrete
representation is formally ill-defined for non-contiguous regions of the
data, including holes and edges of the finite field.  The sharp fall-off
of the convolution kernel implies that only neighboring regions will
be affected.  Again one can account for
these problems by assigning to the unmeasured or contaminated 
regions zero signal
but a substantially larger noise variance than either the
signal or noise in the neighboring measured region.  The mapping
procedure then  propagates an appropriately large and correlated
noise into the reconstruction.   If the gaps in the data are comparable
to the contiguous regions, then Wiener filtering should serve
as the primary mapping technique.

In the limit of infinitesimal pixels and an infinite contiguous field 
both these subtleties disappear and
\begin{equation}
\proj{\gamma\kappa}^t \proj{\gamma\kappa} \rightarrow \iden\,.
\end{equation}
So if the noise in the shear is also uncorrelated and 
statistically homogeneous,
$\cnoise{\gamma\gamma} \propto  \iden$ and Eqn.~(\ref{eqn:estimator})
becomes
\begin{equation}
\esignal{\kappa} \rightarrow \proj{\gamma\kappa}^t \data{\gamma}\,,
\end{equation}
which is the discrete form of the Kaiser \& Squires \cite{KaiSqu93}
result.  
Furthermore, the noise in the $\kappa$ reconstruction
$\cnoise{\kappa\kappa} \rightarrow \cnoise{\gamma\gamma}$.  
We will use this approximation for illustration purposes.
These limiting behaviors and their mapmaking implications are
simplest to derive in the Fourier domain (e.g. \cite{Sel98}).

\subsection{Multiple Source Planes}
\label{sec:sourceplane}

So far we have implicitly assumed a single source plane for the
lensing observables.  
In reality the source galaxies will be broadly distributed around
some median set by the depth of the survey.  It is this fact that
makes three-dimensional mapping possible.

For definiteness we will typically assume a median 
redshift $z_{\rm med}=1$ and a functional form \cite{Kai98}
\begin{equation}
\frac{ dN }{d z} = A \frac{dD}{dz} D \exp[-(D/D_*)^4]\,,
\end{equation}
where $D$ is the comoving distance in the fiducial cosmological
model, and $D_*$ is set to reproduce the median redshift.  
The normalization constant $A$ is chosen to match the number
density of faint galaxies on the sky.  We will take 
$\bar n = 3.6 \times 10^5$ deg$^{-2}$ and $\gamma_{\rm rms}=0.3$
for illustration purposes; 
this represents an estimate of the usable galaxies and
the shear noise per galaxy measured
from a space-based platform (A.~Refregier, private communication).

To separate the source galaxies into redshift bins, galaxy redshifts
with errors that are smaller than the bin size are needed.  
Without spectroscopic redshifts, the precision will be limited
by photometric techniques.  We will take a minimum redshift
bin of $\Delta z = 0.025$ to test the potential of future surveys. 
Current surveys return photometric redshifts with $\Delta z \approx 0.06$ 
\cite{Witetal01}.
 We shall see that reconstruction noise
due to the finite number of galaxies per bin 
dominates before this resolution is reached so that the photometric
redshift errors of even current surveys 
are unlikely to be the limiting source of error.

The mapmaking technique for multiple source planes
with independent noise, as is appropriate for the intrinsic ellipticity noise
of Eqn.~(\ref{eqn:shearnoise}),
is the trivial generalization of a single source plane. 
With correlated noise in the shear from systematic
effects or intrinsic galaxy alignments, one forms a data vector of all the observations and 
applies the same  mapmaking algorithm to estimate $\kappa$ in 
source redshift bins with appropriately correlated noise.

\section{Three-Dimensional Mapping}
\label{sec:los}

Given  two-dimensional convergence maps in multiple source planes, it
is in principle possible to reconstruct full three-dimensional density
maps (e.g.\ \cite{Tay01}).
We shall see that in practice true mapping requires a prohibitively
high signal-to-noise ratio in the lensing observables for reasons 
fundamental to the lensing projection.  We focus
here on the radial reconstruction in a single angular pixel 
since the full three-dimensional distribution may be constructed as
a collection of such reconstructions.

\subsection{Radial Mapmaking}
\label{sec:algorithm}

We now take as the data vector $\data{\kappa}$ in $n_z$
source redshift bins (in a given angular pixel) and assume that its
noise properties $\cnoise{\kappa\kappa}$ are defined
by the reconstruction in \S~\ref{sec:observables}.
The model for the convergence $\kappa$ is a radial projection of the
three-dimensional density distribution (e.g.\ \cite{BarSch01})
\begin{equation}
\kappa(z_s) = \frac{3}{2} H_0^2 \Omega_m \int_{0}^{z_s} dz\,
\frac{dD}{dz}\,\frac{(D_s - D) D}{D_s}\,{\dela}\,,
\label{eqn:kappaint}
\end{equation}
where $D$ is the comoving distance in a flat universe, subscript $s$
denotes evaluation at the redshift of the source $z_s$, 
and $\dela =\delta/a$ is the density fluctuation with the growth rate
in a matter-dominated universe scaled out.
Discretizing Eqn.~(\ref{eqn:kappaint}) in redshift bins returns the
general equation of mapmaking (see Eqn.~(\ref{eqn:mapmaking})) 
\begin{equation}
\data{\kappa} = \proj{\kappa \dela} \signal{\dela} + \noise{\kappa}\,,
\end{equation}
with
\begin{eqnarray}
[\proj{\kappa\dela}]_{ij} = 
\begin{cases}
\frac{3}{2} H_0^2 \Omega_m \dD_j \frac{(D_{i+1} - D_j)D_j}{D_{i+1}}  
   & \text{ $D_{i+1} > D_j$ } ,\\
 0 & \text{ $D_{i+1} \le D_j$ },
\end{cases}
\label{eqn:denproj}
\end{eqnarray}
where $\dD_j$ is the width of bin $j$, the distances are measured to
the center of the bins, and we have offset the source redshift bins
by 1 so that the projection matrix is purely lower triangular.
For notational simplicity we have assumed that the binning in the
signal and data space is the same, but the
generalization is straightforward.

The reconstructed density field is then given by the general mapmaking
equation
\begin{equation}
\esignal{\Delta} = \weight{\Delta\kappa} \data{\kappa}\,,
\end{equation}
where 
\begin{eqnarray}
\weight{\Delta\kappa} &=& 
\cnoise{\Delta\Delta}
\proj{\kappa\Delta}^t \cnoise{\kappa\kappa}^{-1}\,.
\label{eqn:radialweight}
\end{eqnarray}
Here
\begin{equation}
\cnoise{\Delta\Delta} = 
[\proj{\kappa\Delta}^t \cnoise{\kappa\kappa}^{-1} \proj{\kappa\Delta}]^{-1}\,,
\label{eqn:noisedelta}
\end{equation}
is the noise covariance of the estimator.

As noted by Taylor \cite{Tay01}, the reconstruction is in principle
well-posed and does not require regularization if the density field is
to be recovered to the same redshift resolution and range as the
convergence data.  Our more general treatment accounts for
inhomogeneities and correlation in the noise, and even gaps in the 
data (see \S~\ref{sec:observables}).  More importantly, it
returns the noise covariance of the estimator.
We shall see in the next section
that without knowledge of the noise {\it co}variance
the reconstructed density field cannot be used for any practical purpose.

The multipixel generalization of radial mapmaking concatenates
the vectors for each pixel.   For uncorrelated
noise in $\kappa$-pixels, the result is simply the application 
of mapmaking pixel-by-pixel since the noise matrix is 
then block diagonal in
the pixels.  For correlated noise,
the noise matrices in the multipixel
generalization of Eqns.~(\ref{eqn:radialweight}) and 
(\ref{eqn:noisedelta}) couple neighboring pixels.

\subsection{Noise Properties}
\label{sec:mapdiscussion}

To get a feel for the properties of the reconstruction, consider
an idealization with redshift bins that are equal in comoving width
$\dD$, and noise in $\kappa$ that is both uncorrelated and
homogeneous. Then
\begin{eqnarray}
\proj{\kappa\Delta} &=& C {\bf M} \,,\nonumber\\
\cnoise{\kappa\kappa} &=& \sigma^2 \iden\,,
\end{eqnarray}
where
\begin{equation}
C= \frac{3}{2} (H_0 \dD)^2 \Omega_m\,,
\end{equation}
and
\begin{equation}
[{\bf M}]_{ij} = 
\begin{cases}
(i+1-j) \frac{2j-1}{2i+1} & \text{ $j<i+1$ }\,, \\
0 & \text{ else }\,.
\end{cases}
\end{equation}
With these simplifications the reconstruction matrix is
\begin{equation}
[\weight{\Delta\kappa}]_{ij} = \frac{1}{C}\,\frac{2j-1}{2i-1} \times
\begin{cases}
 1 & \text{ $j=i-2,i$ }\,, \\
-2 & \text{ $j=i-1$   }\,, \\
 0 & \text{ else      }\,. 
\end{cases}
\end{equation}
Note that the projection matrix is lower triangular, and the
reconstruction matrix is tridiagonal.

For $i \gg 1$ the reconstructed density field is essentially the finite
difference approximation of the second derivative of the
convergence data; this is the same second derivative seen in
Taylor's 
continuous method
\cite{Tay01}. 
To understand this result, consider the response in $\kappa$ to
a density fluctuation in a single redshift bin
(also see Fig.~\ref{fig:point}b).
As we move out in redshift, the $\kappa$ response is zero until
we reach the density fluctuation.  As we cross the fluctuation
$\kappa$ undergoes a sudden ``acceleration,'' and thereafter
grows slowly.  With perfect data, one would identify density
fluctuations as regions where the second derivative of $\kappa$
is large.  The problem is that the $\kappa$ response accelerates
from zero and hence will be hidden by noise locally.  This is
the fundamental limitation of weak lensing tomography with real
data.

Taking finite differences of the data
amplifies the noise and strongly correlates it between neighboring
pixels.  For $i \gg 1$ the noise matrix is
\begin{equation}
[\cnoise{\Delta\Delta}]_{ij} = \frac{\sigma^2}{C^2} \times
\begin{cases}
1 & \text{ $j=i-2,\, i+2$}\,, \\
-4 &\text{ $j=i-1,\, i+1$}\,, \\
6  &\text{ $j=i$         }\,, \\
0  &\text{ else          }\,.
\end{cases}
\label{eqn:noisecov}
\end{equation}
The noise covariance has a very particular form, which corresponds
to a finite-difference approximation of a fourth derivative
(see Fig.~\ref{fig:radialmap}b).
That the noise is correlated in this specific way will turn out to
be crucial in extracting any information from the reconstruction.

To see this, consider a toy model where the galaxies are equally
distributed among $n_{z}$ redshift bins that extend to a
cosmologically interesting distance 
$\Omega_m^{1/2} H_0 D_{\rm max} = 1$.
Then $\sigma^2 = n_{z} \gamma_{\rm rms}^2 / n_{\rm gal}$ and
$\Omega_m^{1/2} H_0 \dD = n_{z}^{-1}$, and the
rms noise per bin in the reconstruction  is
\begin{equation}
\sqrt{6} \frac{\sigma}{C} \approx 
1.5 \left(\frac{n_z}{20} \right)^{5/2}
    \left(\frac{n_{\rm gal}}{3.6 \times 10^5}\right)^{-1/2}
    \left(\frac{\gamma_{\rm rms}}{0.3}\right)\,,
\end{equation}			
where we have scaled the result with numbers from \S~\ref{sec:sourceplane}
for degree scale pixels.   
Even with these generous assumptions, the signal-to-noise per bin
in the reconstruction is generally small and only approaches unity
for density fluctuations that approach unity when averaged over
these large volumes.
 
Notice that the noise per bin scales as $n_{z}^{5/2}$, which seems
to suggest that increasing the radial resolution decreases the total
signal-to-noise ratio.  But that would be true only if the noise were
uncorrelated.  Instead, the fact that the noise covariance has a very
specific form that is not seen in realistic signals allows it to be
filtered out.  The blind reconstruction of primary mapmaking is
therefore useful only as a first step in the process.  To extract
information out of the map, one must regularize the reconstruction
with a prior assumption about the signal that is to be recovered.

\section{Large-Scale Structure}
\label{sec:lss}

For scales greater than about $10'$ and the depths reached by
modern surveys, the convergence field in a
typical region of sky is dominated by large-scale structures in the
underlying dark matter density field \cite{weaklss}.  Moreover, the
fluctuations are
in the linear to quasilinear regime where theoretical modeling can be 
expected to give a good prior assumption about the statistics of the
field (\S~\ref{sec:signal}).  
In this limit, one can quantify and better represent the
information contained in the noisy three-dimensional density map obtained from
the primary mapmaking of the previous section.  
We apply two well-known techniques: Wiener filtering
(\S~\ref{sec:wiener}) to represent the map itself, and
the Karhunen-Loeve transform (\S~\ref{sec:KL}) whose eigenmodes
encapsulate and expose the underlying information 
contained in the map.

\subsection{Signal Matrix}
\label{sec:signal}

In the linear regime, the signal matrix $\csignal{\dela\dela}$
is related to the linear
power spectrum as follows.
The average density fluctuation within the $i$th redshift window
$W_i({\bf x})$ becomes 
\begin{equation}
\Delta_i = a_i^{-1} \int d^3 x\, W_i ({\bf x}) \delta({\bf x})\,,
\end{equation} 
where $\delta({\bf x})$ is the density fluctuation field.
We assume that the windows are normalized so that 
$\int d^3 x\, W_i =1$.
The signal covariance of these density averages is
\begin{eqnarray}
\left[\csignal{\Delta\Delta}\right]_{ij} & = &
 \int d^3 x_i \int d^3 x_j\, W_i ({\bf x}_i) W_j({\bf x}_j) \nonumber\\
&& \quad \times \langle \delta({\bf x}_i) \delta({\bf x}_j) \rangle 
\\
&=& \frac{G_i}{a_i} \frac{G_j}{a_j} 
\int \frac{d^3 k}{(2\pi)^3} W_i({\bf k}) W_j^*({\bf k}) P(k)\,,\nonumber
\label{eqn:variancesimp}
\end{eqnarray}
where $P(k)$ is the linear power spectrum today, $G_i = G(z_i)$ is the 
linear growth rate of the density field and
$W_i({\bf k})$ are the Fourier transforms of the windows.
Note that in a matter-dominated universe $G(z) = a$ and so
$\Delta$ then has the interpretation of the density field extrapolated 
to the present in linear theory.

For definiteness, we will take the windows to be 
a series of slices in redshift at comoving 
distance $D_i$ and
width $\dD_i$ with a sky pixel radius $\Theta_s= \sqrt{A/\pi}$ in radians
in the small angle approximation and a flat spatial geometry:
\begin{equation}
W_i({\bf k}) = 2 e^{i k_\parallel D_i} 
	\frac{\sin k_\parallel \dD_i/2 }{
        k_\parallel \dD_i/2 }  
        \frac{ J_1(k_\perp D_i \Theta_s) }{
        k_\perp D_i \Theta_s}\,.
        \label{eqn:pillwindow}
\end{equation}

For pixels smaller than $A \sim 1$ deg$^{2}$, the density field
is in the mildly non-linear regime 
\cite{JaiSel97}.  Here the signal matrix should
be calculated with a numerical simulation of structure but a simpler
approximation here suffices to extract the rough scaling.  We replace
in Eqn.~(\ref{eqn:variancesimp}) 
\begin{equation}
G_i G_j P(k) \rightarrow [P_{\rm NL}(k,z_i) P_{\rm NL}(k,z_j)]^{1/2}\,,
\end{equation}
where $P_{\rm NL}(k)$ obtained from $P(k)$ through the
scaling relations of \cite{PeaDod96}. This approximation says that
structures maintain the same coherence as in linear theory
across redshift bins. 

\subsection{Gaussian Realization}
\label{sec:gaussian}

The signal matrix can then be used to make a Gaussian simulation of 
structure.
Consider the Cholesky decomposition of the signal matrix
\begin{equation}
\csignal{\dela\dela} = \chol{\rm S} \chol{\rm S}^t\,,
\end{equation}
and a vector of independent Gaussian random numbers of unit
variance ${\bf g}$, i.e.\ $\langle {\bf g} {\bf g}^t \rangle={\bf I}$.
Then
\begin{equation}
\signal{\dela} = \chol{\rm S} {\bf g}
\end{equation}
is a Gaussian realization of the correlated signal vector since
\begin{equation}
\langle \signal{\dela} \signal{\dela}^t \rangle = 
\chol{\rm S}  \langle {\bf g} {\bf g}^t \rangle \chol{\rm S}^t = 
\csignal{\dela\dela}\,.
\end{equation}
Again, below the degree scale, the Gaussian approximation begins
to break down.  However as tested in simulations it remains
a reasonable approximation down to $10'$ \cite{WhiHu00}.

\begin{figure}[tb]
\centerline{\epsfxsize=3.5truein\epsffile{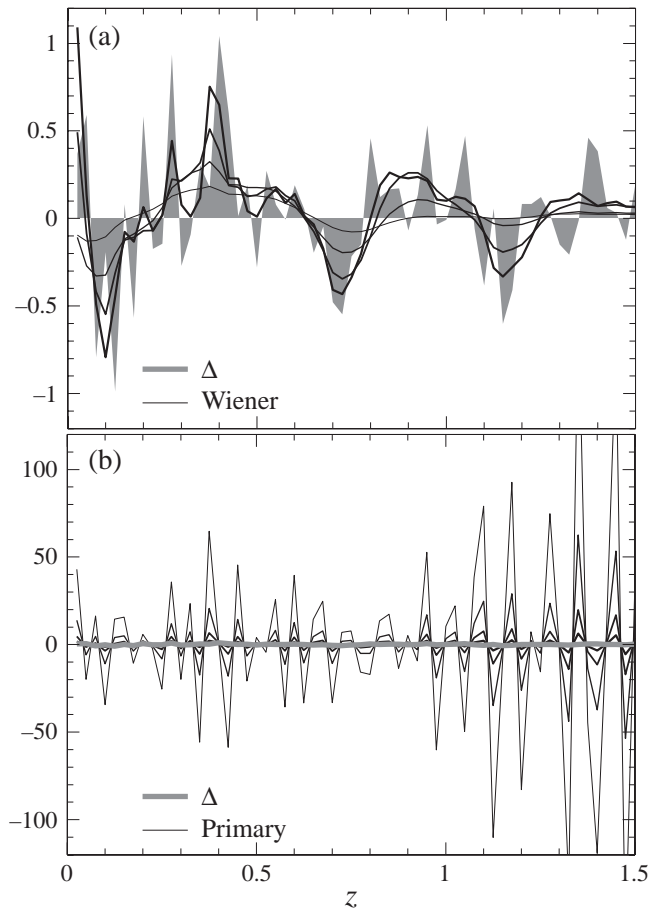}}
\caption{True radial density distribution (shaded) vs.\ 
(a) Wiener and (b) primary reconstructions, for a $4$ deg$^2$ pixel and
source redshift bins with $\Delta z = 0.025$ extending to
$z_{\rm max} = 2.5$.
Thin lines correspond to the fiducial noise variance
($\bar n = 3.6 \times 10^{5}$ deg$^{-2}$ and $\gamma_{\rm rms}=0.3$),
and lines of increasing thickness correspond to noise variance a
factor of 10, 100, and 1000 times lower.}
\label{fig:radialmap}
\end{figure}

We show a sample realization in Fig.~\ref{fig:radialmap}a 
(shaded) for a $\Lambda$CDM cosmology with 
with parameters
$\Omega_c = 0.3$, $\Omega_b=0.05$, $\Omega_\Lambda=0.65$, $h=0.65$,
$n=1$, $\delta_H=4.2\times 10^{-5}$ ($\sigma_8=0.92$), pixel area of
4 deg$^{2}$ and redshift binning of $\Delta z = 0.025$.  
We will use this as the fiducial cosmology in the examples that follow. 

\subsection{Wiener Filter}
\label{sec:wiener}

As discussed in \S~\ref{sec:mapmaking}, Wiener filtering minimizes the
reconstruction noise using prior knowledge of $\csignal{\dela\dela}$, the
covariance matrix of the signal.
In Fig.~\ref{fig:radialmap}, we compare the Wiener reconstruction with
the primary map for a Gaussian realization of structure and several
choices of the noise variance.
The thinnest lines (smoothest for Wiener, noisiest for primary)
correspond to the fiducial noise specifications 
($\bar n = 3.6 \times 10^{5}$ deg$^{-2}$ and $\gamma_{\rm rms}=0.3$;
see \S~\ref{sec:sourceplane}),
and the noise variance decreases by factors of 10 as the lines thicken.
Note the hundredfold difference in noise scale between the Wiener
and primary reconstructions.

The primary reconstruction is far too
noisy to recover a visual impression of the structure, even for
wildly optimistic assumptions about the noise.  However, the noise
has a specific oscillatory structure arising from the noise
covariance (see Eqn.~(\ref{eqn:noisecov})), which is neither
completely random nor present in the true signal.  The Wiener filter
uses the information in the noise covariance of the primary
reconstruction to reveal the hidden signal.
For the fiducial noise specification, the Wiener filtered map recovers
the low order, long-wavelength features in the density field;
it is not until a prohibitively low noise variance is reached that
fine-scale features of order the bin width are recovered. 

The Wiener reconstruction is useful in cases where a map with well-defined
statistical properties is needed, for example for cross-correlation studies
with luminous tracers of the dark matter.

\subsection{KL Transform}
\label{sec:KL}

In the low
signal-to-noise regime, it is more 
quantitatively useful to express the data in terms of
a new set of orthogonal basis functions that are rank ordered
by their signal-to-noise ratio.  Low signal-to-noise modes may be 
eliminated from the data set allowing a near lossless compression
of the data. The pixel representation of these 
modes then tells us their correspondence to the radial density field.
This is accomplished by the Karhunen-Loeve transform, also known as
the signal-to-noise eigenmode technique (e.g.\ \cite{TegTayHea97}).

Consider the generalized eigenmode problem,
\begin{equation}
\csignal{\dela \dela} \eigen = {\epsilon} \cnoise{\dela \dela}\eigen\,.
\end{equation}
With a Cholesky decomposition
\begin{equation}
\cnoise{\dela \dela} = \chol{\rm N}\chol{\rm N}^t\,,
\end{equation}
the generalized eigenmode problem reduces to an ordinary
one
\begin{equation}
 \left[ \chol{\rm N}^{-1} \csignal{\dela\dela} \chol{\rm N}^{-t} \right]
 [\chol{\rm N}^{t} \eigen] = \epsilon [\chol{\rm N}^{t} \eigen]\,.
\end{equation} 
The eigenvectors represent linear combination of the data  $\data{\dela}$.
If one composes the matrix with rows representing the eigenvectors
\begin{equation}
\weight{{\rm K}\dela} = \left(
\begin{matrix}
\eigen_1^t \\
... \\
\eigen_{n_{z}}^t 
\end{matrix}
\right)\,,
\end{equation}
the new representation of the data vector becomes
\begin{equation}
\esignal{\rm K} = \weight{{\rm K}\dela} \data{\dela}\,.
\end{equation}
The important property of the Karhunen-Loeve transform is that
\begin{eqnarray}
\langle \esignal{{\rm K}} \esignal{{\rm K}}^t \rangle  
&=& \weight{{\rm K}\dela} 
[\csignal{\dela\dela}+\cnoise{\dela\dela}] 
    \weight{{\rm K}\dela}^t\nonumber\\
&\equiv& \csignal{{\rm K}{\rm K}} + \cnoise{{\rm K}{\rm K}}\,,
\label{eqn:klcov}
\end{eqnarray}
where the the covariance matrices satisfy the important
condition
\begin{eqnarray}
\csignal{{\rm K}{\rm K}} &=& \epsilon \iden \,,\nonumber\\
\cnoise{{\rm K}{\rm K}}  &=& \iden \,,
\end{eqnarray}
such that the modes are uncorrelated separately in each.
Furthermore, $\epsilon$ quantifies the relative contributions of
signal and noise in the mode.

\begin{figure}[tb]
\centerline{\epsfxsize=3.5truein\epsffile{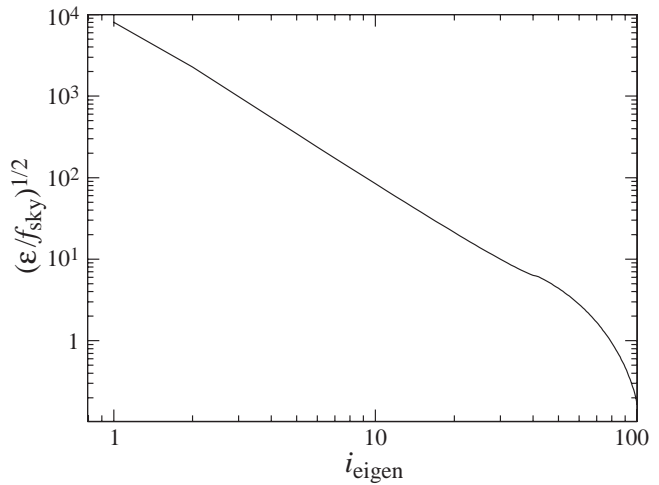}}
\caption{Signal-to-noise ratio in the Karhunen-Loeve eigenmodes with 
the sky-coverage scaled out, $(\epsilon / f_{\rm sky})^{1/2}$,
calculated for $5' \times 5'$ pixels with the fiducial noise variance
and redshift binning.}
\label{fig:snlam}
\end{figure}

\begin{figure}[tb]
\centerline{\epsfxsize=3.5truein\epsffile{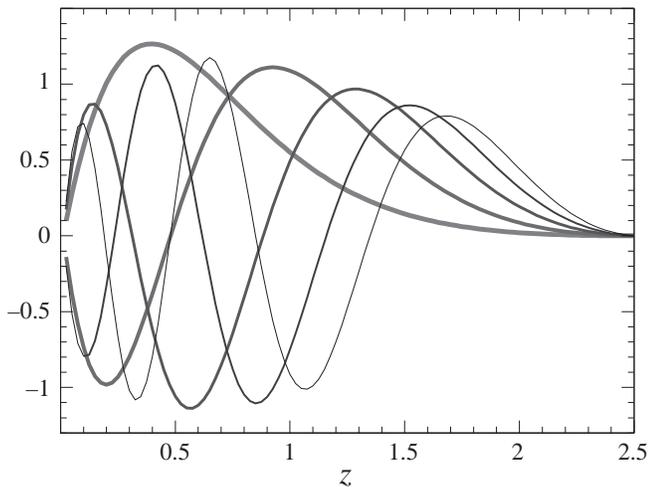}}
\caption{Signal-to-noise eigenmodes for the five largest eigenvalues
in Fig.~\protect{\ref{fig:snlam}}.  The eigenmodes are similar
to low frequency Fourier modes in the radial direction.}
\label{fig:eigen}
\end{figure}

In Fig.~\ref{fig:snlam}, we show the signal-to-noise ratio per
eigenmode $\epsilon^{1/2}$.  This ratio is scaled by the square root
of the fraction $f_{\rm sky}$ of the sky covered by a survey, to
reflect the increase in the signal-to-noise for a statistical
detection with independent pixels.  Here we have assumed a pixel
size of $5' \times 5'$, which is sufficiently small to extract most
of the information on large-scale structure.

Note the steep decrease in the signal-to-noise ratio as a function
of the eigenmode index: the first few eigenmodes contain most of the
information.  This fact allows a radical compression of the
data, here from 100 bins to a handful.
Nonetheless, even though the ratio is small for essentially
all of the higher eigenmodes on the scale of individual pixels 
($f_{\rm sky} \sim 2 \times 10^{-7}$), a survey comprising
4--400 deg$^{2}$ ($f_{\rm sky} \sim 10^{-4}$--$10^{-2}$) has 
more than enough signal for a statistical detection.

To understand the information stored in the higher modes, we plot
the first few eigenmodes in redshift in Fig.~\ref{fig:eigen}
renormalized to have unit norm.
The first eigenvector simply shows the overall lensing efficiency
when integrated over the whole source distribution, i.e.\ it has a
single peak at a distance halfway to the median redshift $z=1$.  
The higher eigenvectors increase the number of nodes with the 
boundary conditions that the weight is negligible near the observer
at $z=0$ and well beyond the median redshift.  They are therefore the
analogues of low order Fourier wavevectors in the radial direction.

These low order modes are sensitive to the cosmological model
itself in that their values depend on the growth rate of 
structure, the volume element of the pixel-redshift bins, and
the distances in the lensing efficiency.  Even a statistical
measure of their rms amplitude can help constrain cosmology, 
in particular the dark energy.

A potential problem for this use of three-dimensional mapping
is that a cosmology is assumed both in the Karhunen-Loeve
decomposition and in the projection matrix itself
Eqn.~(\ref{eqn:denproj}).  The former problem is readily handled
in that even if the a priori assumption of the signal matrix
in Eqn.~(\ref{eqn:variancesimp}) is incorrect, the Karhunen-Loeve
transform is a well-defined linear operation on the data.
The modification comes about in the calculation of the
covariance of the estimators in Eqn.~(\ref{eqn:klcov}).  The
covariance is still diagonal in the noise but need not be 
diagonal in the signal nor do its elements have the interpretation
of signal-to-noise.  Still, it is calculable for the purpose
of model fitting and does not present a fundamental problem.

The second problem is apparently more subtle but reduces
to the same issue.   Errors in our cosmological
assumptions in the projection matrix make the primary map not
correspond precisely to a density reconstruction.  
Fortunately the form of the projection matrix is similar
in all cosmologies: a broad bell-shaped weighting that
peaks halfway to the source distance.
Again, the well-defined linear operations involved allow
us to predict the statistics of the primary map given
a cosmology in spite of the fact that it does not strictly
represent the density field.

Of course, if the recovered cosmology differs greatly from the assumed
one, then the signal-to-noise eigenmodes will become an inefficient
representation of the data.  The best solution to both problems
is to iterate the analysis and converge on a fiducial cosmology
that fits the data.

\section{Individual Dark Matter Halos}
\label{sec:halos}

Below $10'$, the convergence field is dominated by individual
structures, or dark matter halos, along the line-of-sight to the
source galaxies.  The abundance in redshift of such objects
is well known to be exceedingly sensitive to the
growth rate of structure and hence the dark energy in the
universe \cite{abundance}.  Identification and mass measurement
by lensing would be ideal because the association of luminous observables 
with the dark mass
of the halos is always problematic.  Indeed there may exist halos
that are effectively dark \cite{darkcluster}.
However, the efficacy of a 
purely lensing-based study is severely compromised by 
projection effects \cite{projection},
so three-dimensional mapping in principle holds the
key to utilizing this fundamental test.

In the discrete halo case, one also has a well-motivated 
prior to regularize the inversion.  In \ref{sec:mem}, we discuss
a modified version of the maximum entropy method.  This method
is most useful in the intermediate regime where the signal contains
individual objects embedded in the large-scale structure.  We
then describe point source regularization, which is the
best method when there is good reason to believe that effectively
{\it all\/} of the structures are well localized and the large-scale
structure component can be ignored.

\subsection{Maximum Entropy Method}
\label{sec:mem}

The maximum entropy method (MEM) is widely applied in situations 
where a noisy image is assumed to contain both discrete objects and
a diffuse component (e.g.\ \cite{NarNit86}) and has been applied
to two-dimensional weak lensing data \cite{SeiSchBar98}.   
In this case the noisy image is the primary map $\data{\dela}$, the
discrete objects are dark matter halos, and the diffuse component
is the large-scale structure of the universe. MEM involves adding 
a penalty function to the mapmaking 
minimization of Eqn.~(\ref{eqn:minimize}) for the recovery of an 
MEM filtered signal $\signal{{\rm E}}$,
\begin{equation}
H = \lambda  \sum_{i=1}^{n_z} I[\signal{{\rm E}}]_i \ln 
  I[\signal{{\rm E}}]_i\,,
\end{equation}
where the intensity $I$ is some functional of 
$\signal{\rm E}$ that we require 
to be positive.  The Lagrange multiplier
$\lambda$ trades off between minimizing $\chi^2$ and regularizing the 
solution.  It is here chosen to give $\chi^2 = \nu$, the degrees of
freedom.  The main feature of MEM is that
while it prefers a uniform solution $I_i =$const., it does not
additionally penalize bin-to-bin fluctuations as polynomial 
regularization would.   Hence it allows solutions with discrete
objects that occupy only one bin.  

To apply MEM to our mass reconstruction we need to choose $I$.
A natural choice would be $1 + a_i [\signal{\rm E}]_i$ since the density
cannot fluctuate to negative values.  However this prescription
would still strongly disfavor placing all of the mass in a single
redshift bin.  To allow us more freedom in the regularization
let us take a more general form
\begin{eqnarray}
I[x] &=& 1+x \,, \nonumber\\
  x  &=& \delta_{\rm thr} \arctan 
		\left( \frac{a_i [\signal{\rm E}]_i}{\delta_{\rm thr}}\right)\,.
\end{eqnarray}
The parameter $\delta_{\rm thr}$ places a threshold density above
which MEM no longer penalizes the reconstruction; it returns the
natural choice when $\delta_{\rm thr} \rightarrow \infty$.  

\begin{figure}[tb]
\centerline{\epsfxsize=3.5truein\epsffile{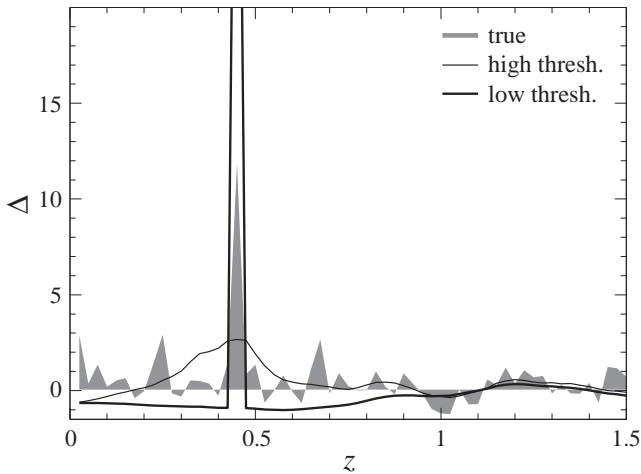}}
\caption{MEM reconstruction with a halo of
$5 \times 10^{14} h^{-1} M_\odot$ added to a $10' \times 10'$ pixel
realization of large-scale structure (shaded) 
with the fiducial noise variance
and redshift binning (see Fig.~\ref{fig:radialmap}).
High (thick line, $\delta_{\rm thr}=10$) and
low (thin line, $\delta_{\rm thr}=3$)
threshold cases are shown.  The
MEM solution makes a discrete transition between these stable solutions.}
\label{fig:mem}
\end{figure}

In Fig.~\ref{fig:mem} we show an example of the technique for
the case of a halo embedded in large scale structure.  For a high
$\delta_{\rm thr}$ compared with the true density contrast of the halo, 
MEM seeks to regularize the reconstructed density
in pixels and returns a smooth solution.  For a $\delta_{\rm thr}$ that
is lower, MEM places a density spike at the right position but the
wrong amplitude.  It favors a solution where the neighboring bins 
are all underdense since there is no penalty for further adding to
the height of the spike.   

A fundamental drawback of MEM is the difficulty in assessing 
the errors in the reconstruction, i.e.\ the reality of the discrete
objects MEM finds.
Note that by construction both solutions in Fig.~\ref{fig:mem}
have exactly the same $\chi^2$, and the radical change in character
of the solution is driven by the prior assumption of 
$\delta_{\rm thr}$ which sets the likelihood of having a comparable
density spike in the solution.
Still, MEM can identify interesting regions in the data for further
study, perhaps with the point source method below.
Conversely, it provides a useful cross check on the robustness
of the point source solutions below.

\subsection{Point Source Method}

In regions that are known to be atypical of large-scale
structure --- either as flagged by the MEM reconstruction or
simply because the signal in $\kappa$ is much too large to be
generated by large-scale structure --- it is reasonable to
assume as a prior that the density field is dominated by a
collection of discrete massive objects.  Ironically, this form
of anti-regularization of the density field in redshift bins is
itself the most extreme regularization of the ones considered
here, i.e.\ it has the least number of allowed degrees of freedom.  
The single-object form of
this technique has been applied to data by \cite{Witetal01} and
yields impressively {\it precise\/} predictions of the redshift or
radial location.  Whether the predictions are {\it accurate},
however, depends on the validity of the single-object assumption
and on the regularization criteria more generally.

Let us state the criteria in a more general form.  Define the
penalty function on a point-source--regularized reconstruction
$\signal{\rm H}$ as 
\begin{equation}
H = 
\begin{cases}
  0     & n_{\rm O} \le n_{\rm H}\,, \\
\infty  & n_{\rm O}  >  n_{\rm H}\,,
\end{cases}
\end{equation}
where $n_{\rm O}$ is the number of redshift bins occupied by 
a density fluctuation, and $n_{\rm H}$ is a prior assumption of the
number of discrete objects (halos) the reconstruction should have.
In other words, the minimization of Eqn.~(\ref{eqn:minimize})
is strictly over the position and density amplitude of $n_{\rm H}$
objects.  
The danger in this method of course is that it will return a best
fit for the $n_{\rm H}$ objects even if the solution is in fact a
smooth distribution or composed of some other number of objects.

A well-defined procedure that makes minimal use of prior information
is to identify sky pixels like to contain one or more massive objects
(as described above), and to perform a sequence of minimizations with
$n_{\rm H} = 1, 2, \ldots$, stopping when the $\chi^2$ does not improve
significantly.

\begin{figure}[tb]
\centerline{\epsfxsize=3.5truein\epsffile{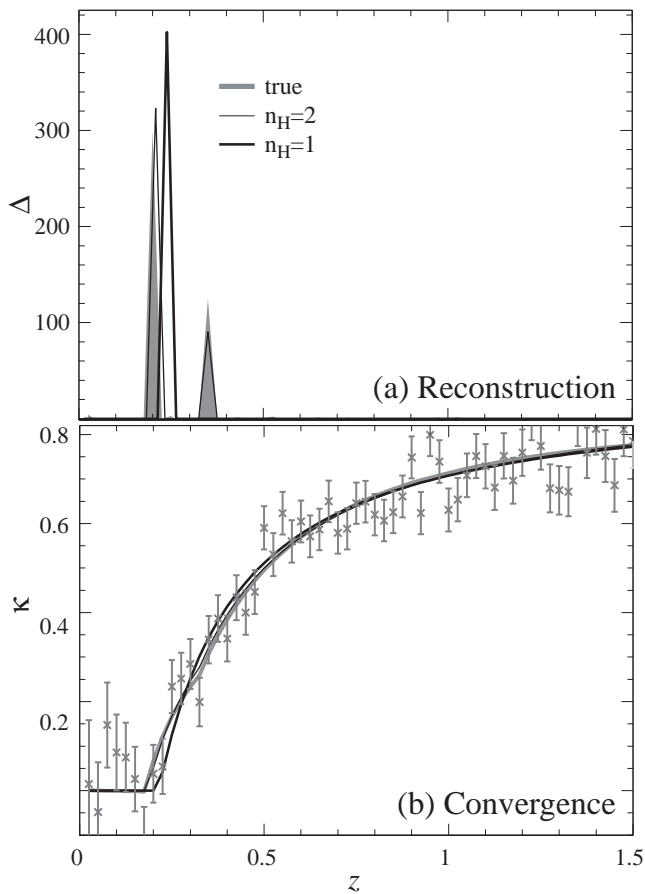}}
\caption{Point source method.
(a) Density reconstruction for two halos of mass
$10^{15} h^{-1} M_\odot$ each added to a $5' \times 5'$ pixel
realization of large-scale structure (shaded), with the fiducial noise variance
and binning (see Fig.~\ref{fig:radialmap}).  Priors of $n_{\rm H}=1$ (thin)
and $n_{\rm H}=2$ halos (thick) are compared.  
(b) Original data in the convergence $\kappa$ compared with the 
reconstructions.}
\label{fig:point}
\end{figure} 

In Fig.~\ref{fig:point}a, we show an example with two very massive
cluster-sized halos at $z=0.19$ and $z=0.34$. 
(The masses are enclosed entirely within the $5' \times 5'$ pixels
and their respective redshift bins.)
The fit assuming one point source returns
$\chi^2 = 96$ for 100 redshift bins and two parameters --- a perfectly
good fit.  The fit yields a redshift constraint $z = 0.239 \pm 0.006$
that is remarkably precise, but wrong.
Going to $n_{\rm H}=2$ does recover two objects in the proper locations,
with $\chi^2 = 91$ for two fewer degrees of freedom.
In Fig.~\ref{fig:point}b we show the implied $\kappa$ fields plotted
against the original data.  The residuals for the one-object fit show
coherent structure near the true halo locations and so the improvement
in $\chi^2$ is significant.
Still, this example warns against blindly interpreting the formal errors
of the fit.  It is actually an optimistic example because the large
masses and redshift separation $(\Delta z=0.15)$ yield a signal 
much larger and much better separated
than expected for real weak lensing measurements.

To better distinguish between close alternatives one could 
fold in prior information.
For example, one could use the theoretically well-understood
abundance of massive halos 
to determine the relative likelihood
of the solutions given the recovered masses of the objects, e.g.\
two halos of $5 \times 10^{15}$ $h^{-1}\, M_\odot$
may be favored over one halo of $10^{16}$ $h^{-1}\, M_\odot$ due
to the predicted exponential suppression in the number
density of high mass halos.  One could also use model profiles 
to create matched filters across smaller pixels that resolve the
halo. 
Finally, prior information from
photometric redshifts of galaxies likely to be members of the 
cluster(s) could decide between competing solutions \cite{Witetal01}.

\section{Discussion}
\label{sec:discussion}

We have shown that the evolution of the shear field
in source redshift contains large-scale information about the distribution
of dark matter, statistical information about its fluctuations
on smaller scales, and redshift localization information for massive
dark matter halos on the smallest scales where the signal is large.    
This information is hidden in the noise of a direct reconstruction, 
and its extraction requires mild prior assumptions about the statistical
properties of the density field.  
We have argued for an approach that begins
with a lossless direct or primary reconstruction that is followed
by regularization by a prior that is appropriate for the information
that is to be extracted.

In the large to intermediate scale regime, the information content can be
distilled into signal-to-noise or KL eigenmodes which efficiently
compress the data by a factor of 10 or more.  These low-order modes probe
the slow evolution of the statistics of
the density field and are well suited to studying the
properties of the dark energy.  Tomographic sensitivity to the
dark energy has been previously
noted in the two-point correlation of the shear 
through the improvement of projected measures
of the dark energy density and equation of state for future 
surveys \cite{tomopower}.
The KL eigenmode decomposition retains information from the
higher order correlations in the field \cite{skewness}
and also establishes a more direct,
non-parametric quantification of the information
contained in the data.   A full study of the cosmological
implications is beyond the scope of this paper,
but we believe that it will be a promising approach for the
future.   

Wiener filtering in the large-scale regime returns
large scale maps of the density field with well-defined 
statistical properties.  These should be useful in
cross-correlation studies with luminous, biased tracers
of the dark matter such as galaxies and galaxy clusters \cite{bias}.
With information on the radial dimension, information on the
evolution of the bias can be recovered which in turn
constrains the tracers' formation and evolution.

In the individual halo regime, tomographic techniques have already
been successfully applied to data \cite{Witetal01}.  With a well-motivated
prior on the number of discrete halos along the line of sight, 
the reconstruction can yield excellent localization of the object(s),
in principle to a precision that is better than that in the
source redshifts themselves.  However the accuracy is compromised
by an incorrect assumption of the number of objects.   
Here we advocate a combined approach of adding discrete objects
to the fit, regularizing by maximum entropy,
and employing prior information and followup.

While it is unfortunate that these prior assumptions are necessary
for extracting information from three-dimensional reconstructions
of the density field, they are generally well-motivated and
testable.  Gravitational lensing therefore remains our most
direct, assumption-free means of probing the distribution of 
the dark matter.

{\it Acknowledgments:} We thank J.~Frieman and A.~Kravtsov 
for useful conversations. 
WH is supported by NASA NAG5-10840 and the DOE OJI
program.
CRK is supported by NASA through Hubble Fellowship grant
HST-HF-01141.01-A from the Space Telescope Science Institute, which
is operated by the Association of Universities for Research in
Astronomy, Inc., under NASA contract NAS5-26555.

\end{document}